# Soliton interaction in a fiber ring laser

D. Y. Tang, B. Zhao and L. M. Zhao

School of Electrical and Electronic Engineering, Nanyang Technological University,

Singapore

H. Y. Tam

Department of Electrical Engineering, Hong Kong Polytechnic University, Hong Kong

## Abstract

We have experimentally investigated the soliton interaction in a passively mode-locked fiber ring laser and revealed the existence of three types of strong soliton interaction: a global type of soliton interaction caused by the existence of unstable CW components; a local type of soliton interaction mediated through the radiative dispersive waves; and the direct soliton interaction. We found that the appearance of the various soliton operation modes observed in the passively mode locked fiber soliton lasers are the direct consequences of these three types of soliton interaction. The soliton interaction in the laser is further numerically simulated based on a pulse tracing technique. The numerical simulations confirmed the existence of the dispersive wave mediated soliton interaction and the direct soliton interaction. Furthermore, it was shown that the resonant dispersive waves mediated soliton interaction in the laser always has the consequence of causing random irregular relative soliton movement, and the experimentally observed states of bound solitons are caused by the direct soliton interaction. In particular, as the solitons generated in the laser could have a profile with long tails, the direct soliton interaction could extend to a soliton separation that is larger than 5 times of the soliton pulse width.

# 1. Introduction

Due to their intrinsic stability optical solitons have been proposed to use as information carrier for the long-distance fiber optic communications and the optical signal processing. A major limitation for the soliton applications is the soliton interaction. To understand the nature and consequence of soliton interaction, extensive theoretical and experimental investigations have been carried out [1-9]. In the framework of the solitons described by the nonlinear Schrödinger equation (NLSE), two sorts of soliton interaction have been revealed: the direct soliton interaction and the long-range dispersive waves mediated soliton interaction. Gordon has firstly studied the direct soliton interaction in optical fibers [1]. He found that solitons exert forces on their neighbors if they are closely spaced. Either attractive or repulsive interaction can result depending on the relative phase difference between the solitons, and the strength of the interaction decreases exponentially with the soliton separation. Such a direct soliton interaction was first experimentally verified by Mitschke and Mollenauer [2]. Direct soliton interaction can be reduced in transmission links by the use of band-pass filters or fast saturable absorbers [3, 4]. Apart from the direct soliton interaction, Smith and Mollenauer also reported the observation of a long-range soliton interaction caused by the existence of dispersive waves [5]. Compared with the direct soliton interaction, this long-range soliton interaction is soliton phase independent.

A real long-distance soliton fiber optic communication system also involves in the soliton losses and periodic amplifications. Therefore, the dynamics of solitons in the system is not described by the NLSE, which is integrable and describes only the conservative systems, but by the Ginzburg-Landau equation (GLE), where soliton is formed not only as a result of the balanced interaction between the fiber Kerr nonlinearity and dispersion, but also as a result of the balance between fiber losses and gain generated by the fiber amplifiers. As a soliton propagating in the system periodically experiences loss and amplification, dispersive waves with discrete spectra are generated, which resonantly draw energy from the soliton. For a soliton to survive in the system, the optical amplifier gain must also balance this dispersive wave emission loss of the soliton. Socci and Romagnoli have analyzed the soliton interaction induced by these resonant dispersive waves [6]. They have shown that the resonant dispersive waves could result in a long-range soliton interaction as well as enhance the direct soliton interaction. Either quasi-bound states of solitons or random wiggling of the soliton positions could be caused by the resonant dispersive wave mediated soliton interaction. Afanasjev and Akhmediev [7], Malomed [8, 9] have also theoretically investigated interactions between the solitons described by the GLE. They have found that bound states of solitons with fixed, discrete soliton separations could be formed under the direct interaction between the solitons.

In certain sense a passively mode-locked soliton fiber ring laser mimics a miniature of a soliton fiber optic communication system. Study on the soliton propagation and interaction in the laser cavity therefore gives a direct insight into the soliton interactions in the long-haul optic communication systems. So far soliton operation of passively mode-locked fiber lasers has been extensively investigated [10-15]. Various modes of soliton operation such as soliton bunching [10], stable randomly spaced soliton distribution [11], harmonic mode locking [12], and transient stochastic soliton evolutions have been experimentally observed. Recently, a state of twin-pulse solitons has also been experimentally revealed in the laser [13]. Theoretical studies on the possible physical mechanisms for the formation of the various soliton operation modes have also been carried out, e.g. Kutz *et. al.* have theoretically investigated the effect of laser gain depletion and recovery on the soliton distribution in the lasers [14], Pilipetskii *et. al.* studied the fiber acoustic effect on the soliton interaction in a fiber soliton laser [15]. There is no doubt that these effects always exist in a fiber soliton laser and affect the soliton dynamics. However, how strong are their influences? Do they play a dominating role on the soliton dynamics in a laser? Does any other soliton interaction exist, which plays an even stronger role on the soliton operation of a laser? And how are these different soliton operation modes formed in a laser? All these questions have so far not been clearly addressed. We believe to clarify them is not only important for understanding the soliton operation of the passively mode-locked fiber soliton lasers, but also potentially useful for the future application of solitons in the ultra-high-bit-rate optical communication systems.

We have conducted detailed experimental and numerical studies on the soliton interaction in a passively mode-locked fiber soliton ring laser. In this paper we present results of our studies. The paper is organized as following: section 2 provides results of the experimental investigations. Based on our experimental results, we have identified the existence of three different types of soliton interaction that played a dominating role on the soliton dynamics in the laser: a global type of soliton interaction caused by the existence of unstable continuous wave (CW) emission; a local type of soliton interaction caused by the dispersive waves emitted by the solitons; and the direct soliton interaction. We show experimentally that each of the three types of soliton interaction has different interaction range and strength, and depending on the concrete laser operation conditions they can exist either individually or concurrently. In section 3 we present the numerical simulations on the soliton interaction in the laser. Based on a coupled GLE model and also considering the effects of dispersive wave emission and laser cavity feedback, we found that the major experimental observations could be well numerically reproduced. In addition, based on the results of numerical simulations we found that the effect of the dispersive wave mediated soliton interaction in the laser is always to cause the random irregular relative soliton movements, and the states of bound solitons observed experimentally are formed through the direct soliton interaction. Section 4 is the conclusion of the paper.

## 2. Experimental setup and results

The fiber soliton laser we used for our experimental studies is schematically shown in Fig. 1. It has a ring cavity of about 5.5 meter long, which comprises of a 3.5 m 2000 ppm erbium-doped fiber (EDF) whose group velocity dispersion is about -10 ps/kn.nm, a piece of 1 meter long single-mode dispersion-shifted fiber whose group velocity dispersion is about -2 ps/km.nm, and another piece of 1 meter standard telecom fiber (SMF28). The nonlinear polarization rotation technique is used to achieve the self-started mode locking of the laser. For this purpose a polarization dependent isolator together with two polarization controllers, one consists of two quarter-wave plates and the other two quarter-wave plates and one half-wave plate, is used to adjust the polarization of light in the cavity. The polarization dependent isolator and the polarization controllers are mounted on a 7 cm long fiber bench to easily and accurately control the polarization of the light. The laser is pumped by a pigtailed InGaAsP semiconductor diode of wavelength 1480 nm. The output of the fiber laser is taken via a 10% fiber coupler and analyzed with an optical spectrum analyzer (ANDO AQ6317B) and a commercial optical autocorrelator (Inrad 5-14-LDA). A 50 GHz wide bandwidth sampling oscilloscope (Agilent 86100A) and a 25 GHz photo-detector (New Focus 1414FC) are used to study the soliton evolution in the laser cavity.

We are interested in finding out the physical mechanisms for the formation of the various modes of the multiple soliton operation, in particular, how they are related to the soliton interaction in the laser. To this end we have designed various experiments to identify and separate the various natures of soliton interactions in the laser. In all our experiments we have simultaneously monitored the laser output by using an optical spectrum analyzer, a 50GHz high-speed sampling oscilloscope, and a commercial autocorrelator. Through carefully analyzing the experimental results, we could clearly identify the existence of three different types of strong soliton interaction in our laser. These are a global type of soliton interaction caused by the existence of unstable CW components, a local type of soliton interaction mediated through the dispersive waves and the direct soliton interaction. Figure 2 shows for example a series of the experimentally measured oscilloscope traces that demonstrate the existence of the global type of soliton interaction in the laser. The solitons observed in our laser have an average pulse width (FWHM) of about 340fs, and the cavity round trip time of our laser is about 26 ns. In the state shown in Fig. 2, there are five solitons coexisting in the cavity. Two solitons have a very small soliton separation (~ 6 ps) and are bound together. Details about the properties and binding mechanism of this bound-soliton will be discussed later. As our detection system cannot resolve these two solitons, it appears as a large pulse on the oscilloscope trace. We used the bound-soliton as the trigger signal for our measurements (it therefore has a fixed position in the oscilloscope traces), and experimentally investigated the relative movement of the other three solitons in the cavity. Figure 2a and 2b show the relative position changes of the three soliton-pulses under fixed laser parameters. We observed that the three solitons ceaselessly moved with respect to each other with an average speed of about half a division/second. Their movements were purely random. The solitons could come close or move apart from each other. In such a state of laser operation, the movement of one soliton affects all the other solitons in the cavity no matter how far apart they are. We therefore believe that there is a global type of soliton interaction between them. We further investigated the physical mechanism of the global soliton interaction and found out that it was caused by the existence of an unstable CW component in the laser. To demonstrate this, we have shown in Fig. 3 the experimentally measured soliton spectrum of the state. It is to see that there is actually CW lasing coexisting with the soliton operation. Through purely adjusting the pump power, the strength of the CW lasing can be controlled. As the laser is operating in the anomalous dispersion region, a CW component is intrinsically unstable due to the modulation instability (MI). We observed that when the strength of the CW component was strong, it became unstable as characterized by the appearance of the modulation instability sidebands in the spectrum [16]. In the oscilloscope trace, instability of the CW component is also represented by the appearance of background noise, as can be clearly seen in Fig. 2a and 2b. Reducing the strength of the CW component, its instability becomes less obvious, and consequently the noise background disappears from the oscilloscope trace as shown in Fig. 2c. Without the noise background all solitons then stay stable in the cavity, forming a stable irregular distribution pattern.

Coexistence of CW lasing and soliton operation was frequently observed in passively mode-locked fiber soliton lasers [16-18]. Interaction between solitons in the presence of a frequency-shifted CW wave was also investigated by Loh *et. al.* [19]. It was found that the existence of a weak frequency-shifted CW caused the soliton central frequency to shift. So far we have not fully understood how the solitons interact with an unstable CW in the laser cavity. However, considering that both the CW component and the solitons coexist in the laser cavity and share the same laser gain, the existence of mutual influence between them seems easy understandable. Hence we may explain the experimental observations in the following way. When a stable CW wave coexists in the cavity, all the solitons would experience the same central frequency shift, overall the relative soliton velocity will not change. While if the CW component becomes unstable, each soliton would experience different local perturbations, therefore, they would have different central frequency shifts. Passive harmonic mode locking of soliton pulses has been frequently observed in our laser [18]. Experimentally we noticed that all the harmonically mode-locked states were actually obtained under the existence of unstable CW components. This experimental finding suggests that the unstable CW component induced global soliton interaction must have played an important role on the formation of the harmonic mode locking. Intuitively it is to imagine that under the global soliton interaction, solitons will gradually adjust their positions. The only stable state they could eventually reach would be a harmonically mode-locked state as only in such a state all the solitons subject to the same influence caused by an unstable CW.

Obviously in a state like that shown in Fig. 2, the soliton interaction induced by the gain depletion and recovery, the fiber acoustic effect should also exist. However, experimentally we observed that once the noise background was suppressed, the solitons then stayed in the cavity stably. They didn't change their positions until the unstable CW component was created again. With this method we had actually controlled the soliton distribution in the cavity and obtained various stable soliton distribution patterns. This experimental result clearly demonstrated that the effects of soliton interaction caused by the gain competition and/or the fiber acoustics are too weak to be measured in the laser.

In Fig. 2 there are two solitons binding together, forming a state of bound solitons. From the autocorrelation trace we found that the two solitons have a fixed pulse separation of about 6 ps. In our experiment as we increased or decreased the background noise, neither the state of bound solitons was destroyed nor the pulse separation was changed, which indicates that the binding force between the two solitons is stronger than the force of the global soliton interaction induced by the unstable CW. Bound states of solitons with various soliton separations are frequently obtained in the laser. Figure 4 shows for example the autocorrelation trace of another state of bound solitons with a larger soliton separation. Coexistence of bound solitons with different pulse separations has also been observed [20]. Experimentally we found that the larger the pulse separation, the weaker is the binding energy between the solitons, therefore, the easier is the state of bound solitons to be destroyed by the perturbations.

A common feature of these states of bound solitons is that the soliton separations have fixed, discrete values, which are independent of the laser operation conditions. Whenever such a bound state of solitons is obtained, it will have one of the values of the soliton separation. As the soliton separations of the bound solitons are fairly large compared with the soliton pulse width, which in the case of our laser is only about 340 fs, based on the theory of soliton interaction of the NLSE solitons [1], it seems unlikely that these states of bound solitons are formed through the direct soliton interaction, as according to the theory when the separation between the solitons are larger than 5 times of their pulse widths, there is practically no overlap between the soliton profiles. Therefore, the existence of these bound solitons in our laser suggests either that the theory of direct soliton interaction of the NLSE solitons does not apply to the solitons of the laser or there exists an unknown soliton interaction that binds the solitons together. However, this needs to be further studied. We note that such a kind of bound solitons has also been observed by other authors in other passively mode-locked soliton fiber lasers [21].

Apart from the above bound states of solitons, depending on the experimental conditions frequently the state of so-called soliton bunching was also observed in the laser. This state was described as another mode of the multiple soliton operation of the passively mode-locked soliton fiber lasers [10]. For the purpose of comparison, we have shown again in Fig. 5 a typical case of such a state observed in our laser. Figure 5a shows that the soliton bunch moves stably in the laser cavity with the fundamental cavity repetition rate. Figure 5b is the expansion of the bunch. 17 solitons coexist in the bunch with fixed relative soliton separations. Note that the soliton separations in the bunch vary randomly, and the separations between the solitons are also fairly large as compared with those of the bound solitons. It is to point out that although a state of soliton bunch is somehow like a state of bound solitons with large soliton separation, the state is far less stable than the bound solitons. Frequently only slightly changing the experimental conditions e.g. the pump power, or even the environmental perturbations could destroy the state. We also note that the state shown in figure 5 is just one typical case of soliton bunching, in the experiment depending on the concrete experimental conditions, especially depending on the history of the soliton operation as also noticed by Grelu et. al. [22], other forms of soliton bunching, such as the one shown in figure 6 where three solitons bunch in the cavity and the formed bunch coexists with other solitons, can also be observed.

Experimentally we found that if two solitons were spaced closely, even after the suppression of the global type of soliton interaction as discussed above, occasionally they could still oscillate relatively to each other, and their oscillations had no effect on the other solitons in the cavity. While when solitons were far apart, they always have stable relative separations. This experimental result suggests that there must exist a local type of interaction between the solitons. We have experimentally checked the maximum range of the local type of soliton interaction. The range varies from laser to laser depending on the laser cavity design and properties. For lasers with long cavity length, this kind of soliton interaction is very obvious and the interaction could extend to more than 200ps, while for our current laser we could just identify this kind of relative soliton oscillations with our detection systems occasionally. This interaction range reminded us the well-know long-range soliton interaction mediated through dispersive waves firstly discovered by Smith and Mollenauer [5]. Dispersive wave emission is a generic property of solitons under perturbations. Dispersive wave mediated long-range soliton interaction has been extensively investigated for the case of soliton propagation [5, 6]. Socci and Romagnoli have theoretically investigated the property of soliton interaction in periodically amplified fiber links and predicted the existence of quasi bound solitons and oscillation type of soliton interaction [6]. However, to our knowledge, no experimental studies of the effect in the fiber soliton lasers have been reported. Different to the cases of soliton propagation the strength of dispersive waves in a fiber laser can be controlled either through carefully adjusting the pump strength or through changing the linear cavity phase delay bias. Different design of the laser cavity can also either suppress or enhance the resonant dispersive wave generation in a laser. All these features of the dispersive waves in a laser make the soliton interaction in the system more complicated.

Theoretical studies on the interaction between the NLSE solitons have shown that when two solitons have a separation of less than about 5 times of their pulse widths, the direct soliton interaction must be considered [1,4]. Although extensive theoretical studies have been done on the direct soliton interaction, in particular, in the case of solitons described by the GLE, formation of bound solitons under the direct soliton interaction was predicted. Nevertheless, only few experimental studies have focused on the direct soliton interaction in passively mode-locked fiber lasers [23, 24]. Partially because direct soliton interaction depends sensitively on the relative soliton phase difference, in practice due to the unavoidable environmental perturbations the relative soliton phase always varies. Even if the environmental perturbations could be suppressed, the existence of the long-range soliton interaction mediated through the dispersive waves as mentioned above could complicate the dynamics of the direct soliton interaction in a laser.

Although in our experiments we could frequently observe that solitons moved across or collided with one another during a transient process, mostly when a laser parameter was suddenly significantly changed. It doubtlessly shows the existence of direct soliton interaction in the system. However, this process occurs normally very fast. It is difficult to experimentally investigate their interaction dynamics. Nevertheless, under certain special laser operation conditions we could experimentally easily obtain stable states of bound solitons with pulse separation less than 5 times of the soliton pulse width. Figure 7 shows for example the optical spectrum and the corresponding autocorrelation trace of one of the states. The autocorrelation trace shows that two solitons of equal pulse height bind tightly together with a pulse separation of merely 920fs. Obviously direct soliton interaction exists between the two solitons. It is to point out that these bound solitons also exhibit fixed, discrete soliton separations. In addition, as the solitons are so close, from their optical spectra we can clearly identify that the relative phase difference between the two bound solitons is always about $\pi$, as can also be seen in figure 7 there is always a dip in the center of their optical spectra.

Coexistence of multiple such bound solitons was also observed experimentally. Surprisingly, different to the bound solitons described above, which can coexist with the unbound solitons in the cavity, the bound solitons as shown in figure 7 have never experimentally found to coexist with other bound solitons or unbound solitons. The multiple such bound solitons can form various modes of operation, such as bunching of the bound-solitons [13] and passive harmonic mode locking of the bound-solitons [18]. Figure 8 shows for example a case of the laser operation with multiple such bound solitons in the cavity, where 10 bound-solitons coexist with stable irregular separations in the cavity. Here again limited by the resolution of the detection system, the separation between the two solitons within a bound-soliton cannot be resolved, however, it is to note that each spike in the oscillation trace is actually a bound solitons with two solitons binding together as detected from their autocorrelation traces and optical spectra. The bound solitons behave as an entity. They can also interact with each other and form bound bound-solitons [20]. All these features suggest that the bound-solitons formed under the direct soliton interaction have very strong binding energy. It could be treated virtually as a new form of soliton in the laser. We note that C. Paré and P. A. Bélanger had theoretically found that a dispersion-managed system can support an antisymmetric stationary pulse consisting of two close spaced dispersion-managed (DM) solitons with $\pi$-phase shifted. They explained the new stationary pulse as a higher-order mode of the fundamental DM soliton [25]. Recently A. Maruta *et. al.* have also numerically found the existence of stable symmetric(in-phase) and antisymmetric (anti-phase) bisolitons as they called in dispersion-managed systems [26-27]. Their numerical simulations also revealed that the bisolitons have fixed discrete soliton spacings. Indeed, the bound solitons of our laser shown in Fig. 7 exhibit close similarity in property to the antisymmertic bisoliton of Paré and Bélanger, and Maruta et. al. which further supports our conjecture that it could be a new form of soliton in the laser.

## 3. Numerical simulations

To have a better understanding on the soliton interaction in the laser, in particular, to find out the effect of the dispersive waves on the soliton interaction, we have also numerically simulated the multiple soliton operation of our laser. Again our simulations are based on the pulse tracing technique reported previously [28]. The basic idea of the technique is to numerically determine the eigenstate of a laser under a certain laser operation condition by circulating light in the laser cavity to a steady state, and then to compare the steady state result with the experimental observations. Concretely, we start from an arbitrary weak light input and let the light circulate in the laser cavity according to the exact laser cavity configuration. As the nonlinear polarization rotation technique was used to achieve mode locking in our laser, where both the linear and nonlinear light polarization rotation within one cavity round trip play an important role on the laser dynamics, to accurately simulate the light polarization effect, we have used the coupled complex Ginzburg-Landau equations to describe the light propagation in the optical fibers,

$$
\begin{aligned}
\frac{\partial u}{\partial z} &= i\beta u - \delta\frac{\partial u}{\partial t} - \frac{i}{2}\kappa''\frac{\partial^2 u}{\partial t^2} + i\gamma\left(|u|^2 + \frac{2}{3}|v|^2\right)u + \frac{i\gamma}{3}v^2 u^* + \frac{g}{2}u + \frac{g}{2\Omega_g^2}\frac{\partial^2 u}{\partial t^2} \\
\frac{\partial v}{\partial z} &= -i\beta v + \delta\frac{\partial v}{\partial t} - \frac{i}{2}\kappa''\frac{\partial^2 v}{\partial t^2} + i\gamma\left(\frac{2}{3}|u|^2 + |v|^2\right)v + \frac{i\gamma}{3}u^2 v^* + \frac{g}{2}v + \frac{g}{2\Omega_g^2}\frac{\partial^2 v}{\partial t^2}
\end{aligned}
\tag{1}
$$

where $u$ and $v$ are the two normalized slowly varying pulse envelopes along the slow and the fast axes of the optical fiber, respectively. $2\beta = 2\pi\Delta n/\lambda$ is the wave-number difference, and $2\delta = 2\beta\lambda/2\pi c$ is the inverse group-velocity difference. $\kappa''$ is the dispersion parameter. $\gamma$ is the nonlinearity of the fiber. g is the gain coefficient for the erbium doped fiber (EDF) and $\Omega_g$ is the gain bandwidth. For the un-doped fiber, g is taken as zero. The erbium doped fiber (EDF) gain is saturated by the total light energy in the cavity as described by

$$g = g_0 \exp\left(-\frac{\int_{-\infty}^{\infty}\left(|u|^2 + |v|^2\right)dt}{E_S}\right),\tag{2}$$

where $g_0$ is the small signal gain and $E_s$ is the saturation energy.

We simulated the effects of other cavity components, such as the polarization controllers and the polarization dependent isolator, by multiplying their transform matrixes to the light field whenever the light field met them. After the light circulated one round trip in the cavity, we used the final light field as the input for the next round of circulation, until a steady state was reached. If no steady state could be obtained, we simply calculated to 5000 rounds and took the final results for our analysis. It is to note that our simulations are different to the conventional master equation method widely used to simulate the operation of soliton lasers [29,30]. It covers all the results obtained by the master equation. In addition, it also has the following advantages: Since the calculation is made following the pulse propagation in the cavity, pulse evolution within one round trip can be studied with the model. As within each step of calculation the pulse variation is always small, even if the overall change of the pulse within one round trip could be big, the model can still give stable solutions. Therefore, the simulation has no limitation on the pulse change within one round trip, and dynamical process of soliton evolution, such as soliton interaction, soliton generation etc can be investigated. In addition, the effect of discrete cavity components, the influence of the dispersive waves and the different order of cavity components on soliton property are automatically included in the calculation. To make the simulated results comparable with our experimental results, we have used the following parameters for our simulations: the laser cavity length L= 6m, the linear cavity beat length $L_b$=L/4, the effective gain bandwidth $\Delta\Omega_g$ = 20nm, the polarization of the intracavity polarizer to the fast axis of the birefringent fiber R=$\pi$/8, the group velocity dispersion (GVD) of the dispersion shifted fiber, the EDF and the standard single mode fiber are $\kappa''$= -2 ps/nm.km, $\kappa''$= -10 ps/nm.km and $\kappa''$= -18 ps/nm.km, respectively. The normalized gain saturation energy of the erbium doped fiber Es=1000. We have only varied the values of the linear cavity phase delay between the two polarization components within one cavity round trip and the small signal gain coefficient of the fiber, these are to simulate the effects of rotating the wavplates and changing the pump, respectively. By using exactly the same numerical model we have previously successfully simulated other effects of the soliton fiber lasers such as the sideband asymmetry and sub-sideband generation [31], periodic soliton peak power modulation [32].

With appropriate linear cavity phase delay selection, and through increasing the pump power, which in our model corresponds to increasing the small signal gain coefficient, self-started mode locking of the laser can be automatically obtained in our simulations. Exactly like the experimental case, immediately after the mode locking coexistence of multiple soliton pulses in the simulation window are always obtained. Decreasing the pump power soliton can be destroyed one by one from the window, and the soliton operation of the laser exhibits clear pump power hysteresis [33,34]. With already solitons in the cavity, increasing the pump power soliton can also be generated one by one in the cavity. Depending on the laser operation parameter settings, the new solitons can be formed either through the soliton-shaping of dispersive waves [31], of the unstable CW components or through the pulse splitting [35]. All these numerical results are well in agreement with the experimental observations, which demonstrate that our simulation can faithfully reproduce the features of the laser. Here we will focus on the soliton interaction in the laser and the comparison between the numerically calculated results to those of the experimental observations.

To simulate the soliton interaction in the laser, we first input two arbitrary weak light inputs in the cavity and let them evolve into solitons with a desired separation, we then changed the laser operation condition to study their interaction. Fig. 9 shows the case of soliton interaction with a large soliton separation. In the steady state it is to see that both solitons have identical soliton parameters such as the peak power and pulse width. Changing the laser operation conditions, both soliton respond in exactly the same way, showing that except sharing the laser gain they have practically no other interactions. In this case even as we increased the pump strength to just before a new soliton was generated, the relative soliton separation remained constant. We then gradually reduced the separation between the solitons. To a separation of about 12ps we found that as we increased the pump to certain strength, the solitons started to change their relative separation and in the meantime the soliton peak powers started to fluctuate randomly, indicating that there was interaction between them and the interaction is the pump strength related. As compared with the soliton pulse width the solitons are still fairly far apart, we attribute the soliton interaction as caused by the dispersive waves. With soliton separations setting below 8ps, we observed that depending on the pump strength, the solitons either formed a state of bound solitons or oscillated relatively. Figure 10 shows for example a case of the bound states of solitons formed. The solitons evolved to the separation and stabilized there. The interaction between the solitons is obvious. The state is called a state of bound solitons because as we varied the pump strength within a certain range, despite the fact that the average soliton energy increased and there is interaction between the solitons, the soliton separation remained the same. Note that the soliton peak powers fluctuate randomly. If we carefully reduced the pump, the peak power fluctuations could be suppressed and the solitons then had identical profiles as shown in figure 10 b. Fig. 11 shows the case of soliton oscillation and collision. Even solitons are initially in a state of bound solitons, when the pump power is increased to beyond a certain value, which varies with the soliton separations, the bound solitons can break their binding and start to oscillate relatively. If the pump power is further increased, the soliton can even collide with each other as shown in figure 11. Nevertheless, the relative soliton oscillations and collisions can be suppressed by simply reducing the pump strength. We emphasize that like the experimental observations, the separations between the calculated bound solitons shown in figure 10 are obviously larger than 5 times of the soliton pulse widths.

Bound states of solitons with soliton separations less than five times of the soliton pulse width were also obtained. Fig. 12 shows for example one of the states. In this case the direct soliton interaction can obviously not be ignored. Different to the bound states of solitons shown in figure 11a, the bound solitons have no peak power fluctuations, even when we increased the pump power to the point where the bound solitons breaks, and afterwards it jumped to a state of bound solitons with a larger soliton separation. Fig. 12b shows the optical spectrum of the bound soliton, which shows that the phase difference between the bound solitons is also about $\pi$. Numerically we found that different bound solitons have different stability as reflected by the range of pump power change over it they are stable. The closer the solitons are separated, the larger is their binding energy. In addition, the soliton separations of the numerically calculated bound solitons also exhibit discrete values, which is in excellent agreement with the experimental observations. At last we emphasize that the calculated results are independent of the initial conditions of the simulation. By simply increased the pump power we can also obtain multiple solitons in the cavity, and we found their interactions exhibit exactly the same features. Particularly, in these cases states of bound bound-solitons, and the formation of multiple identical bound solitons as observed experimentally were obtained, which we will report separately.

## 4. Discussions

Our numerical simulations have well reproduced the typical effects of the soliton interaction in the laser. Based on results of the experimental observations and numerical simulations, we could now explain the formation of the various multiple soliton operation modes of our laser. A fundamental difference between the solitons formed in a laser and in a fiber propagation system is that the laser solitons also subject to the influence of the laser cavity and laser action, e.g. they are not only a soliton pulse but also a mode-locked pulse in the cavity. The mode locking enforces that all the solitons in cavity have exactly the same central frequency and constant phase differences. Therefore, in the steady state and if there is no other force on the solitons, they should have exactly the same properties. And consequently they remain static relative to each other. We believe this should be why we can observe the soliton bunches and stable irregular soliton distributions in the laser. It is to see that these states are actually not a state of bound solitons as there is no interaction between the solitons.

Solitons coexisting in a laser share the same laser gain, therefore, there is gain competition between them. Previous theoretical studies have also shown that soliton interaction caused by the gain depletion and recovery could affect the soliton distribution in the cavity [14]. However, our experimental results shown in Fig.2 clearly demonstrated that the strength of this soliton interaction is actually even weaker than that caused by the unstable CW waves. Therefore, in the practice its effect on the soliton distribution and dynamics in a laser could be ignored. This is also true for the soliton interaction induced by the fiber acoustic effects, as compared with the other types of soliton interactions existing in the laser this type of soliton interaction is too weak to play a role.

Although the bound solitons calculated from the GLE also exhibit fixed, discrete soliton separations, they are only weakly stable as shown by Malomed [8, 9]. The bound solitons formed in the laser are in contrary very stable. At the first sight it seems that they could be formed by different mechanism. Nevertheless, our numerical simulations confirm that the bound solitons formed in a laser are indeed very stable. To explain this difference we believe that the mode-locking feature of the laser solitons could have contributed to it. In addition, our numerical simulation also showed the existence of bound solitons with soliton separations larger than 5 times of the soliton pulse width. We have numerically checked the soliton profiles of our laser. It revealed that the profiles varied with the laser operation conditions, and due to the strong influence of the cavity properties and cavity components, in some cases it is very different to those of the NLSE and the GLE solitons. Although the 3dB pulse width of the soliton is still narrow, the soliton profile could have a tail of considerable strength extending to a distance as large as about 6ps. Direct soliton interaction is certainly the soliton profile dependent. As the solitons formed in different nonlinear systems are different, there is no surprise that bound states of solitons with soliton separation larger than 5 times of the soliton pulse widths could be obtained in the lasers. Therefore, these bound states of solitons observed are still formed by the direct soliton interaction. These experimental and numerical results also suggest that the criteria for identifying direct soliton interaction for solitons of different nonlinear dynamical systems should be different.

Obviously in the laser the direct soliton interaction always coexists with the dispersive wave mediated soliton interaction. It is therefore difficult to separate their effects. There has speculation arguing that the resonant dispersive wave mediated soliton interaction could synchronize the soliton phases and result in the formation of the bound solitons of discrete separations [36]. However, analyzing our numerical results it seems that the dispersive wave mediated soliton interaction in a laser is only to cause random, irregular relative soliton movements. To further identify the effect of the soliton interaction, we have also numerically simulated lasers with different cavity lengths and cavity dispersion. We found that for lasers of long cavity lengths the soliton sidebands can become very strong, and in the meantime the range of soliton interaction can become very long. In a case soliton interaction range as large as about 60ps has been numerically obtained. At this soliton separation the direct soliton interaction could be safely ignored. However, except the random irregular relative soliton oscillations, no bound states of solitons were ever obtained. Assuming that the effect of the dispersive wave mediated soliton interaction is only to cause random irregular soliton movements we found that all the experimental and numerical results obtained could be well consistently explained. With relatively large soliton separation there is no direct soliton interaction, therefore, solitons can only exhibit relative oscillations. With close soliton separations, depending on the relative strength between the dispersive waves mediated soliton interaction and the direct soliton interaction, solitons could either bind together to form bound states of solitons or oscillate relatively. As the dispersive waves in a laser is the pump strength dependent, carefully control the pump can reduce the strength of the dispersive wave mediated soliton interaction, therefore, eliminate the random peak power fluctuation of the bound solitons.

## 5. Conclusions

In conclusion, we have both experimentally and numerically investigated the soliton interaction in a passively mode-locked fiber soliton ring laser. Our experimental studies revealed the existence of three dominating types of soliton interaction in the laser. They are a global type of soliton interaction caused by unstable CW lasing, a local type of soliton interaction caused by the resonant dispersive waves, and the direct soliton-soliton interaction. The impacts of each of the soliton interactions on the formation of the various modes of soliton operation of the laser were investigated. The global type of soliton interaction has the largest interaction range but the weakest interaction strength among the three types of soliton interaction. This type of soliton interaction can be well suppressed by carefully control the pump strength. Experimentally it was found that this type of soliton interaction played an important role on the formation of the passive harmonic mode locking of the laser. The interaction range of the dispersive wave mediated soliton interaction varies with the laser operation conditions, which is determined by the spatial extension of the resonant dispersive waves. This type of soliton interaction always causes random, irregular relative solitons movement. The direct soliton interaction only occurs when the soliton profiles overlap. Our numerical simulation on the soliton interaction is made based on the coupled Ginzburg-Landau equations. It also takes into account the effects of the resonant dispersive waves and the laser cavity properties. For the first time to our knowledge, the details of the soliton interaction in a laser cavity could be numerically faithfully reproduced. Based on results of the numerical simulations it turned out that the solitons in a laser have an intrinsic tendency of approaching synchronization. We conjecture that it is because of this feature of the solitons that resulted in the formation of the states of soliton bunch. The feature of the laser solitons also enhances the stability of the bound solitons formed by the direct soliton interaction. Therefore, bound states of solitons with fixed, discrete soliton separation could be easily obtained in the laser.

**Figure captions:**

FIG. 1. A schematic of the fiber laser setup. PI: polarization dependent isolator; PC: polarization controller; WDM: wavelength division multiplexer; EDF: erbium-doped fiber; DSF: dispersion-shifted fiber.

FIG. 2. Oscilloscope traces demonstrating unstable CW lasing induced global soliton interaction.  (a) and (b): Two consecutive instants under the existence of noisy background. Solitons change their positions in the cavity; (c): Noisy background is suppressed. Solitons form a stable irregular pattern.

FIG. 3. Optical spectrum of the laser output under the same condition of Fig. 2a and Fig. 2b.  The strong line in the spectrum is the CW emission.

FIG. 4. Autocorrelation trace of another bound solitons obtained. Soliton separation is about 8.4 ps.

FIG. 5. Oscilloscope traces of a soliton bunching state. (a) The bunch repeats with the cavity repetition rate. (b): Detailed structure of the bunch.

FIG. 6. Oscilloscope trace showing coexistence of solitons with bunched solitons.

FIG. 7. Optical spectrum of a state of bound solitons under existence of direct soliton interaction.  The insertion is the corresponding autocorrelation trace.

FIG. 8. Oscilloscope trace of the laser output showing coexistence of multiple identical bound solitons. Note that each pulse in the trace consists of two solitons with a separation of about 930fs.

FIG. 9. Numerically calculated soliton evolution in cavity when the solitons are far apart. The cavity linear phase delay bias $\varphi=1.4\pi$. $g_0=350$.

FIG. 10. A numerically calculated bound state of solitons. The cavity linear phase delay bias $\varphi=1.4\pi$. (a) $g_0=370$. (b) $g_0=365$.

FIG. 11. A numerically calculated state of soliton oscillation and collision. The cavity linear phase delay bias $\varphi=1.4\pi$. $g_0=375$.

FIG. 12. A numerically calculated state of bound solitons with close soliton separation. The cavity linear phase delay bias $\varphi=1.4\pi$. $g_0=320$. (a) Time evolutions of the state. (b) Optical spectral variations.

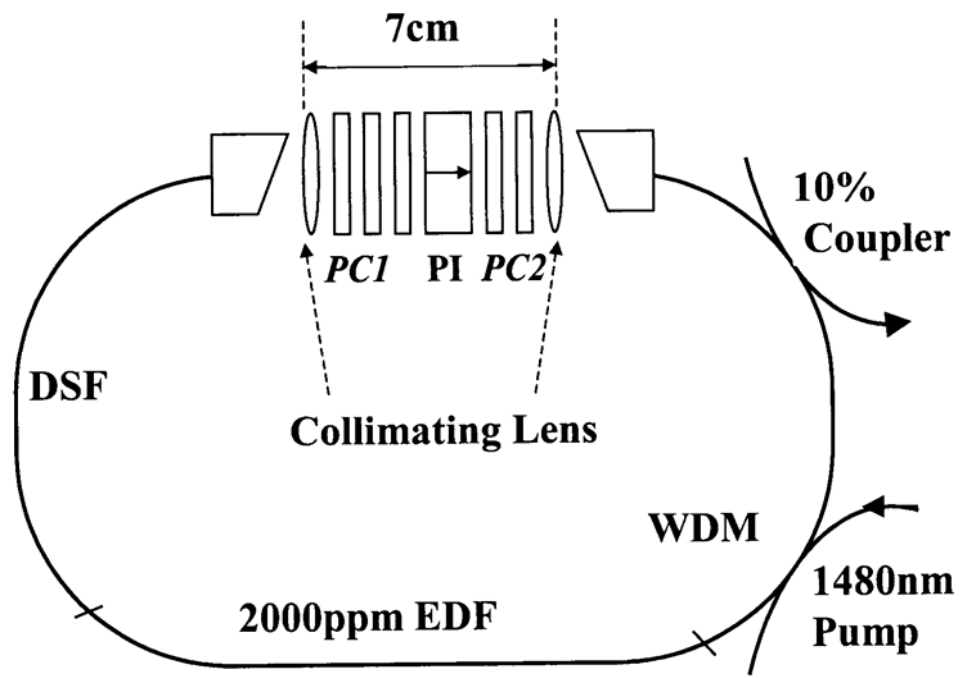

Fig. 1

D. Y. Tang et. al.

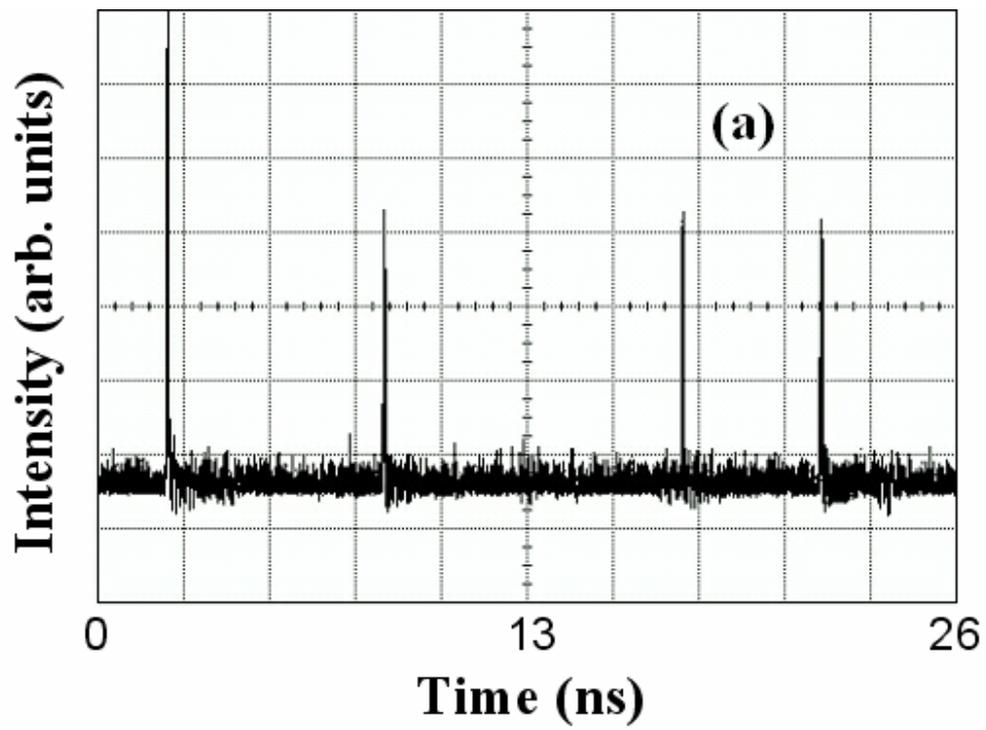

Fig. 2a

D. Y. Tang et. al.

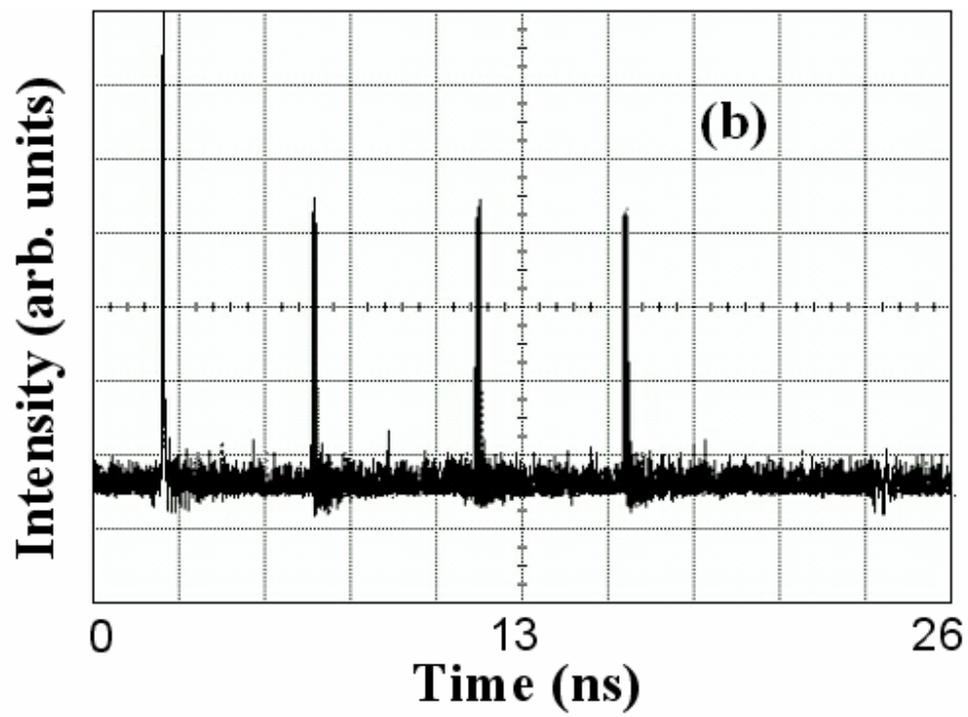

Fig. 2b

D. Y. Tang et. al.

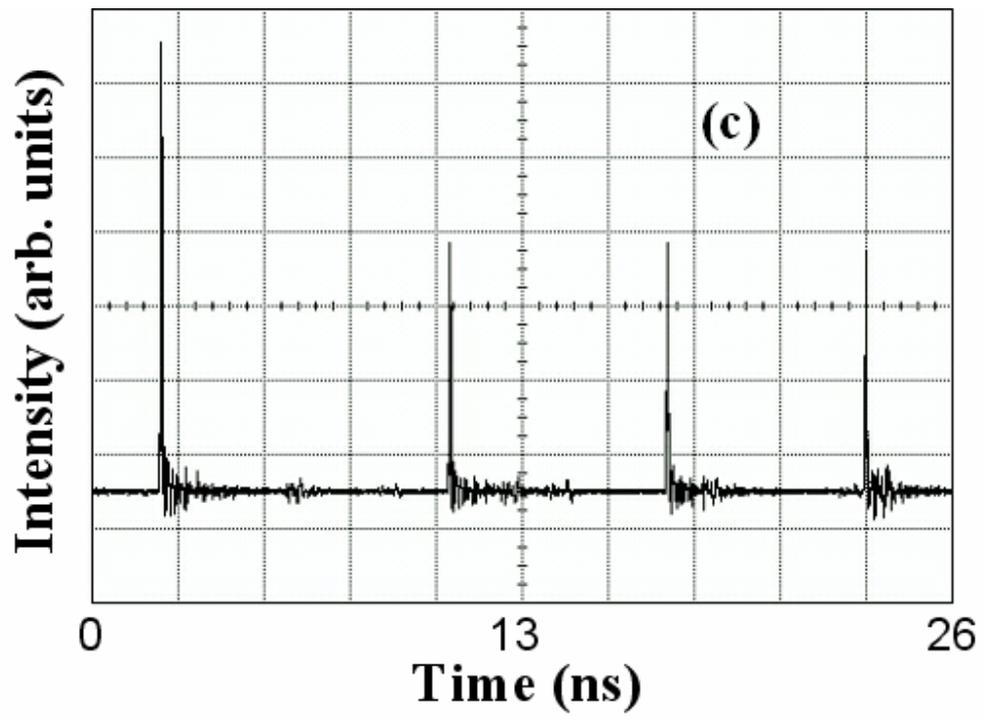

Fig. 2c

D. Y. Tang et. al.

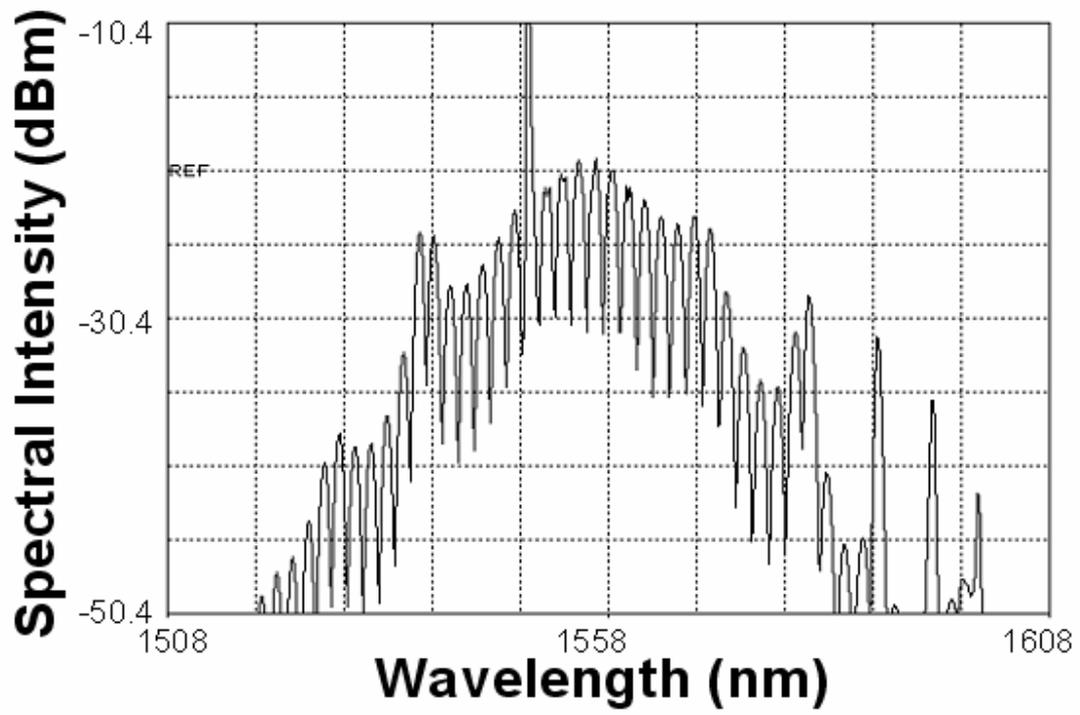

Fig. 3

D. Y. Tang et. al.

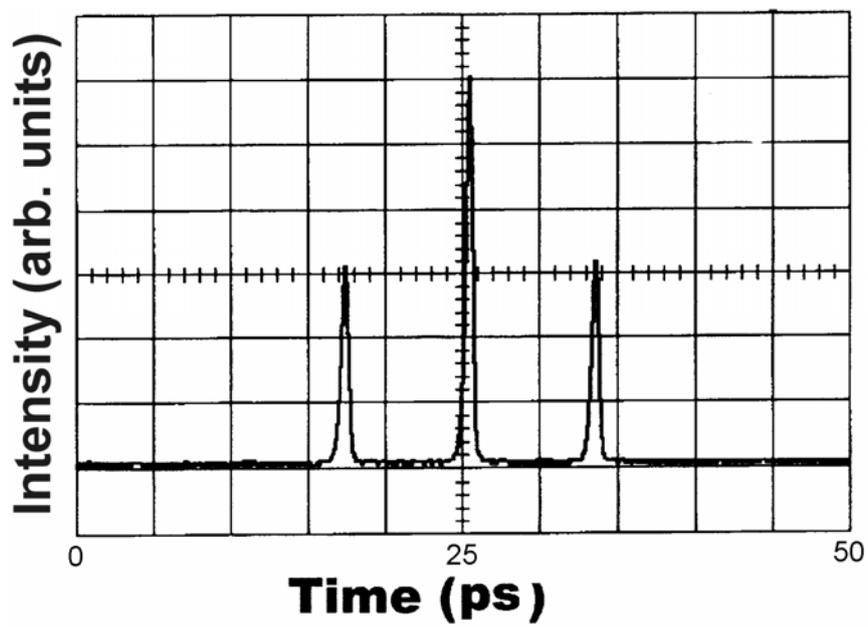

Fig. 4

D. Y. Tang  et.al.

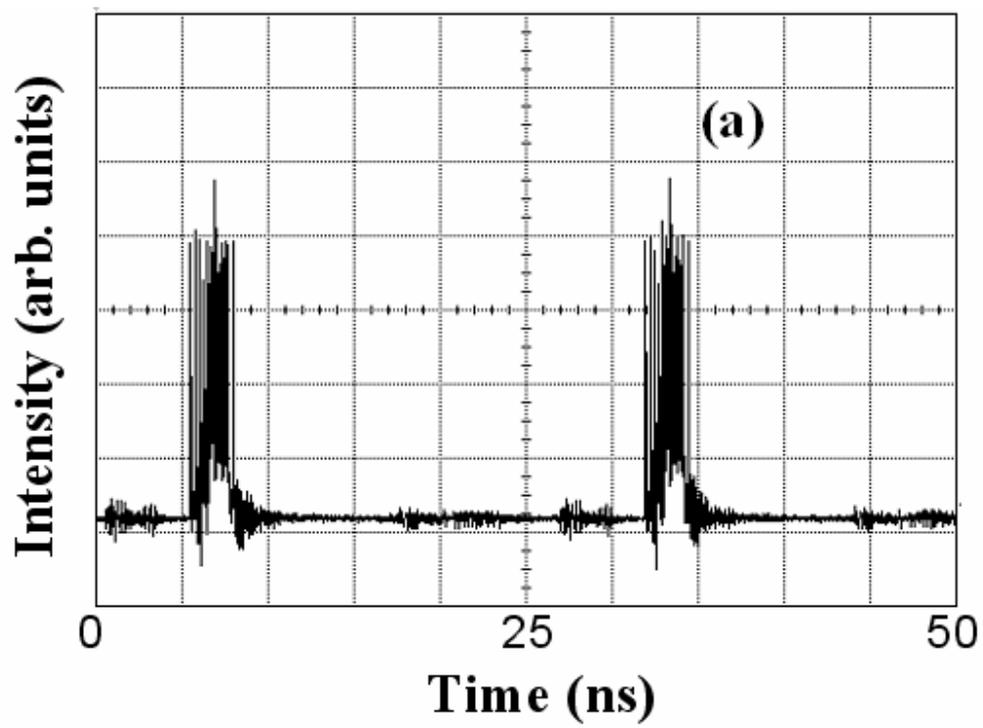

Fig. 5a

D. Y. Tang et. al.

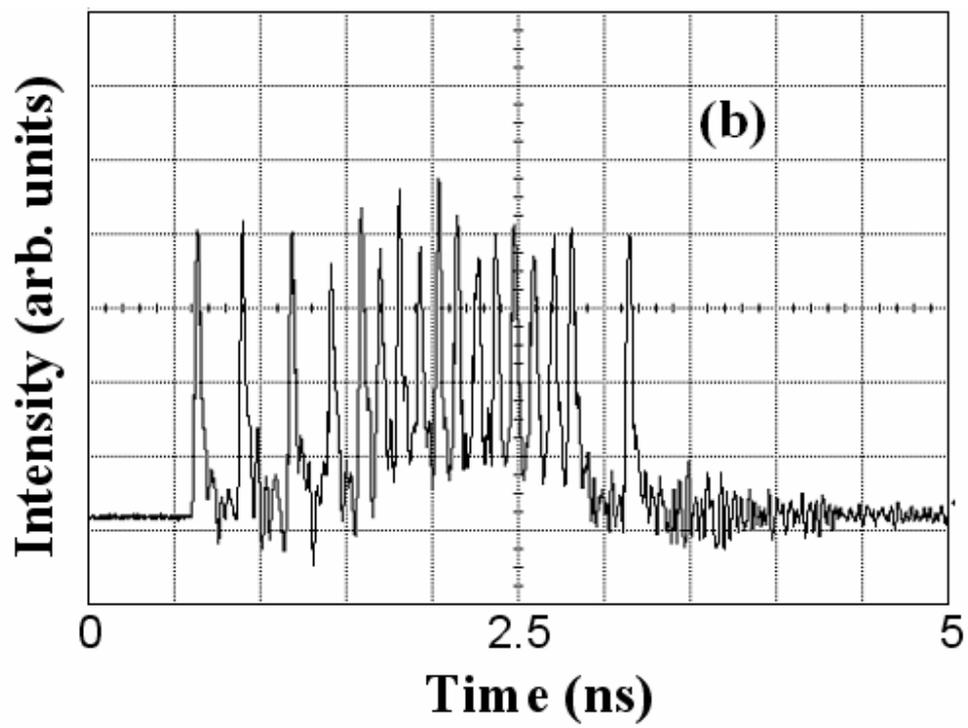

Fig. 5b

D. Y. Tang et. al.

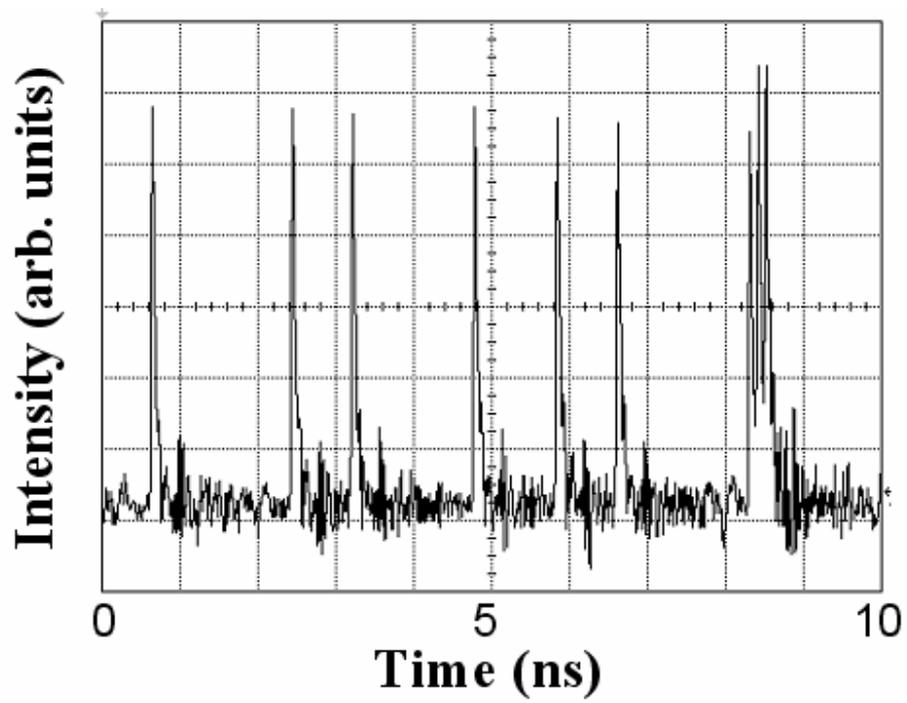

Fig. 6

D. Y. Tang et. al.

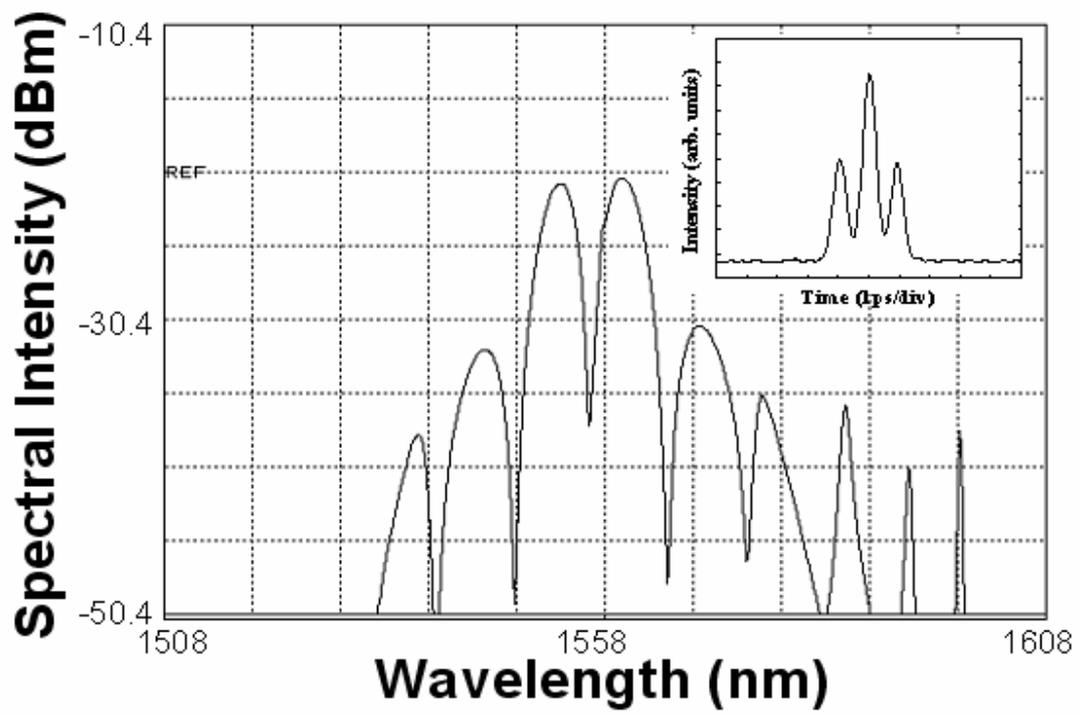

Fig. 7

D. Y. Tang et. al.

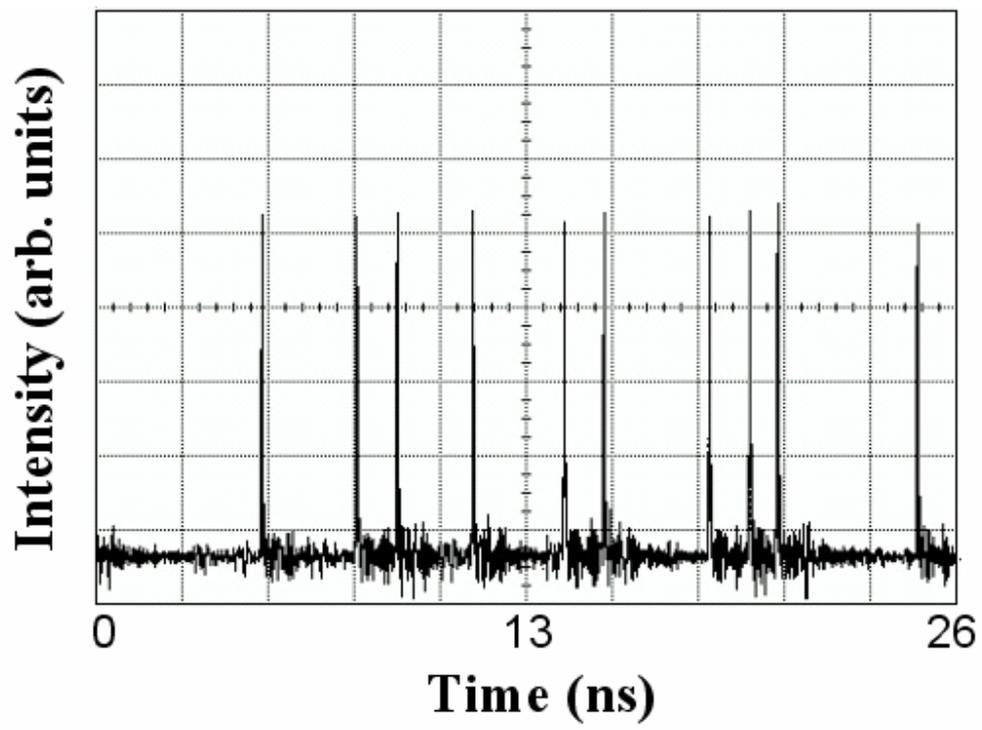

Fig. 8

D. Y. Tang et. al.

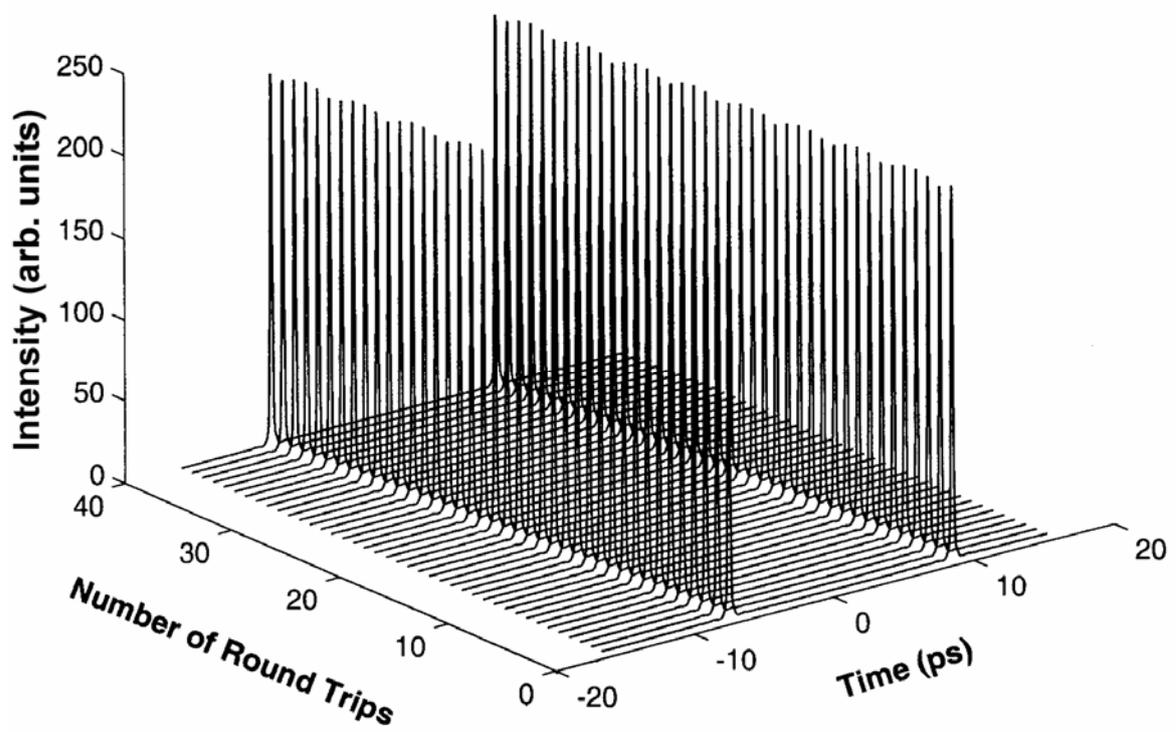

Fig. 9

D. Y. Tang et. al.

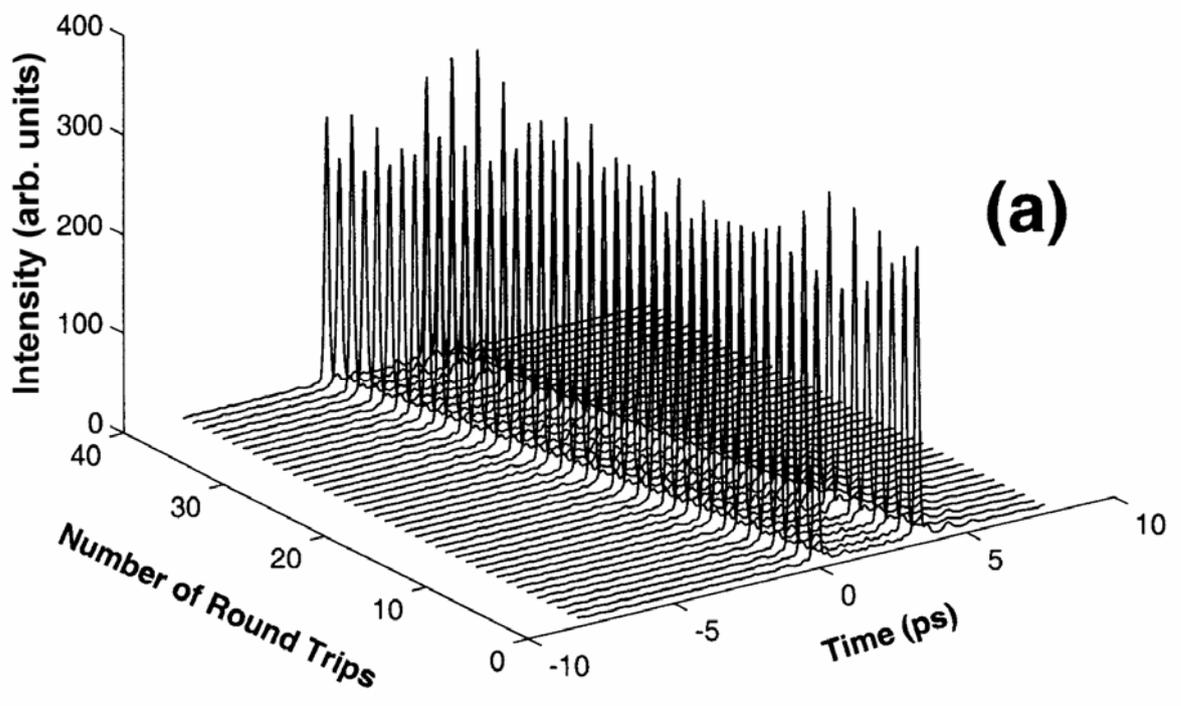

Fig. 10a

D. Y. Tang et. al.

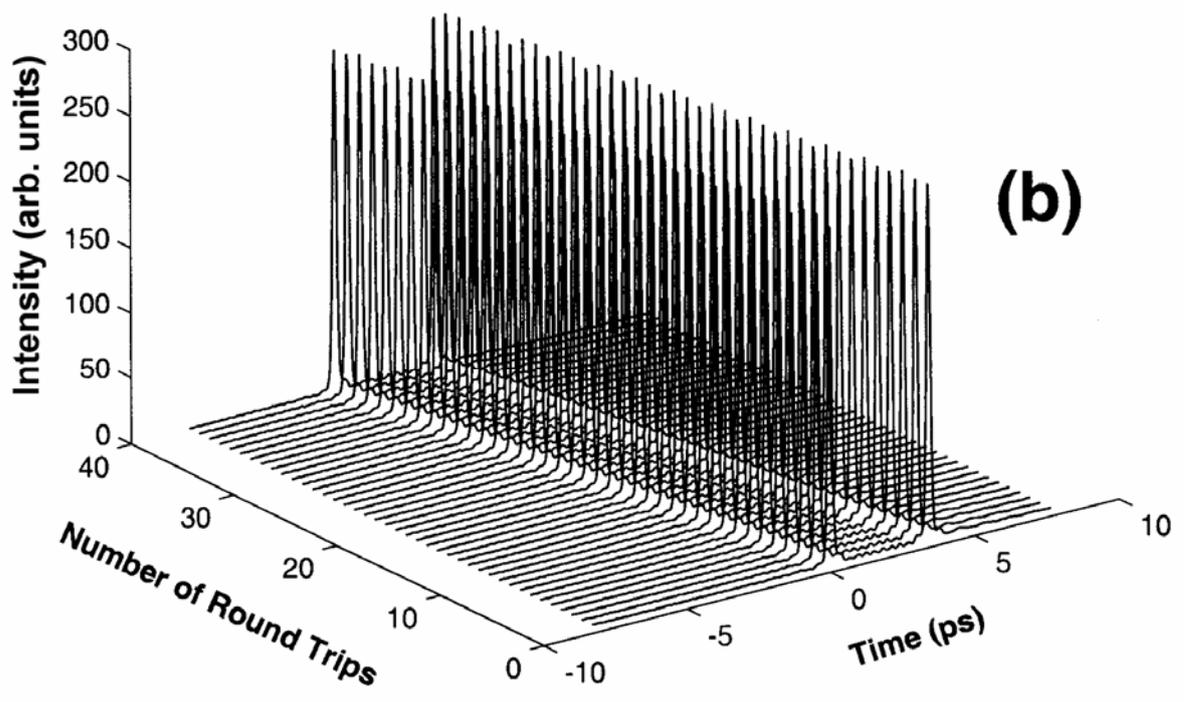

Fig. 10b

D. Y. Tang et. al.

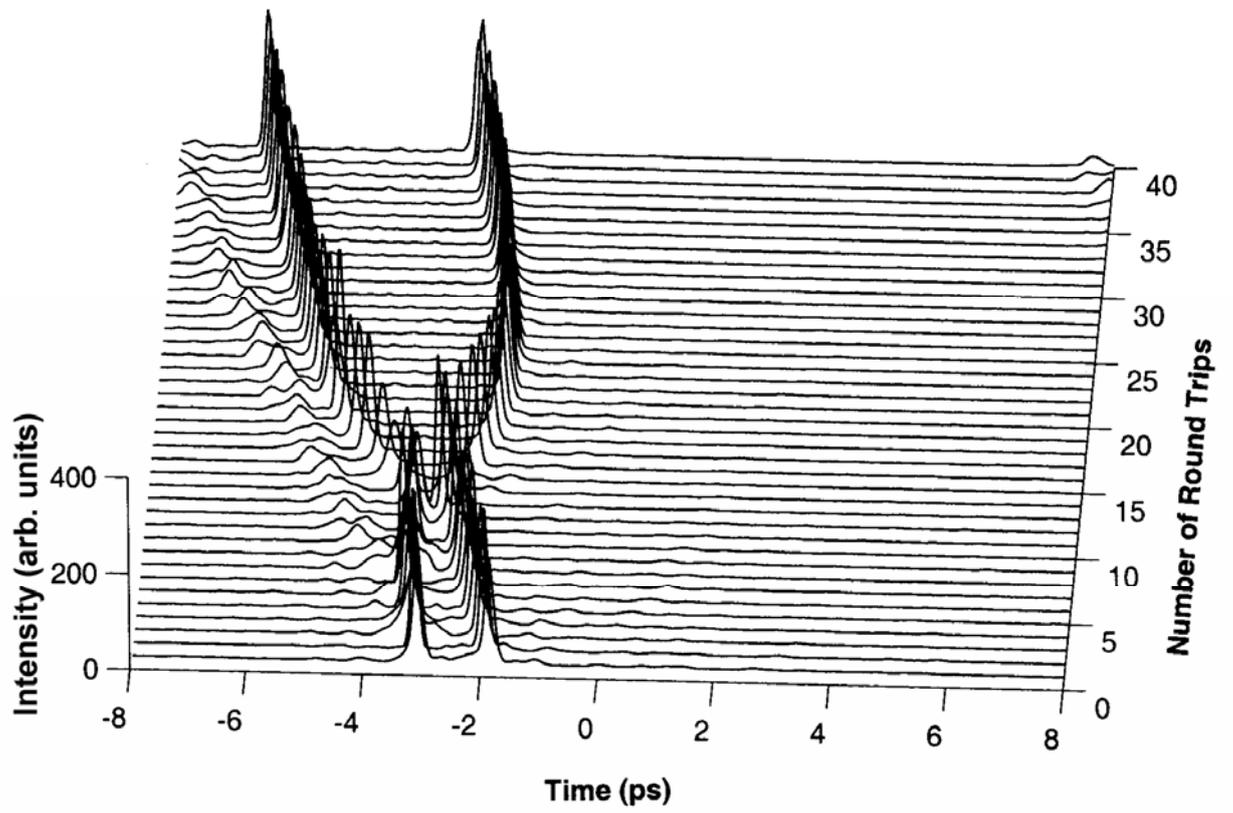

Fig. 11.

D. Y. Tang et. al.

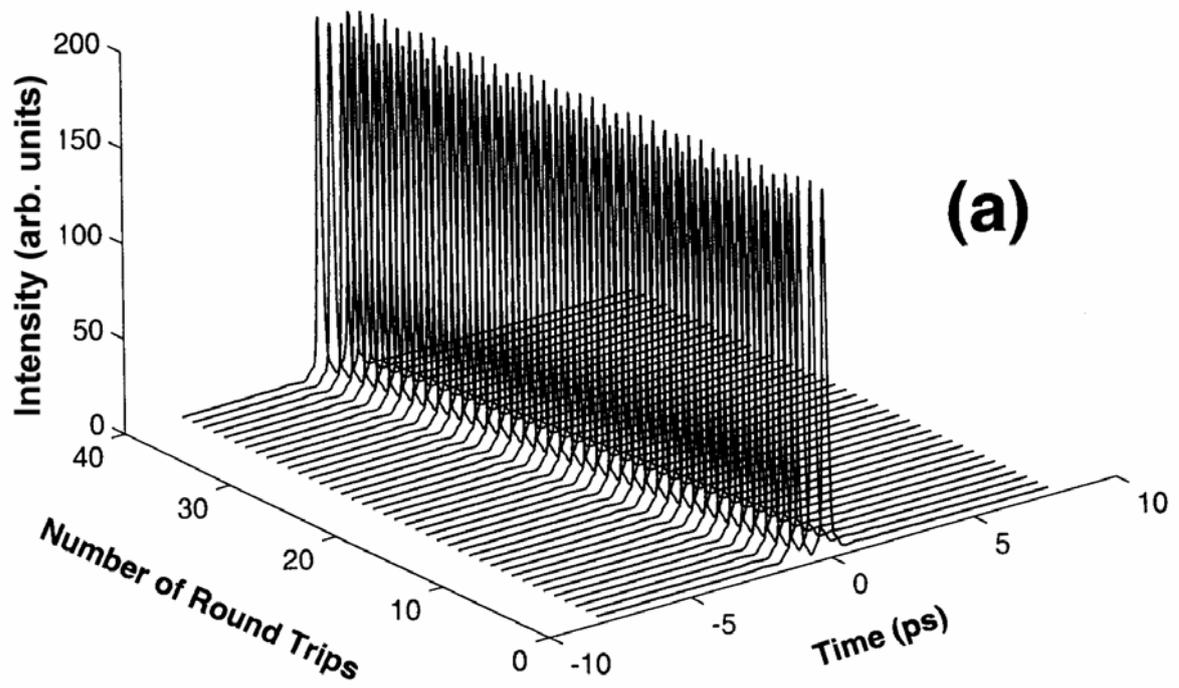

Fig. 12a

D. Y. Tang et. al.

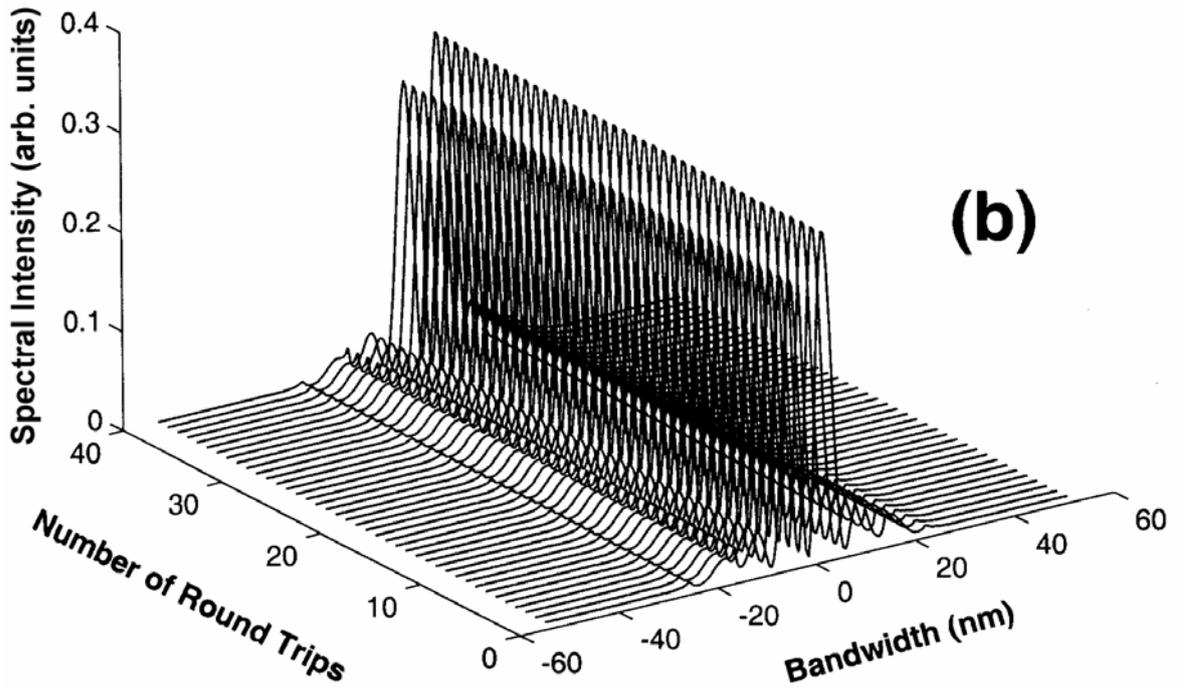

Fig. 12b

D. Y. Tang et. al.